\documentclass{acm_proc_article-sp}

\begin{document}

\title{A New Approach to Coding in Content Based MANETs}
%
%
%
%
%

\numberofauthors{5} 
%
\author{
%
%
\alignauthor
Joshua Joy\\
       \email{jjoy@cs.ucla.edu}
\alignauthor
Yu-Ting Yu\\
       \email{yutingyu@cs.ucla.edu}
\alignauthor
Victor Perez\\
       \email{vperez@cs.ucla.edu}
\and  
\alignauthor
Dennis Lu\\
       \email{slu005@cs.ucla.edu}
\alignauthor
Mario Gerla\\
       \email{gerla@cs.ucla.edu}
}
\date{30 July 1999}

\maketitle

\begin{abstract}
In content-based mobile ad hoc networks (CB-MANETs), random linear network coding (NC) can be used to reliably disseminate large files under intermittent connectivity. Conventional NC involves random unrestricted coding at intermediate nodes. This however is vulnerable to pollution attacks. To avoid attacks, a brute force approach is to restrict the mixing at the source. However, source restricted NC generally reduces the robustness of the code in the face of errors, losses and mobility induced intermittence. CB-MANETs introduce a new option. Caching is common in CB MANETs and a fully reassembled cached file can be viewed as a new source. Thus, NC packets can be mixed at all sources (including the originator and the intermediate caches) yet still providing protection from pollution. The hypothesis we wish to test in this paper is whether in CB-MANETs with sufficient caches of a file, the performance (in terms of robustness) of the restricted coding equals that of unrestricted coding.  

In this paper, we examine and compare unrestricted coding to full cache coding, source only coding, and no coding. As expected, we find that full cache coding remains competitive with unrestricted coding while maintaining full protection against pollution attacks.
\end{abstract}

\section{Introduction}
Traditional Random Linear Network Coding (RNLC) in \\MANETs performs random network coding packet mixing at intermediate nodes. The benefits of RLNC are reliable dissemination of files despite mobility, random interference, and losses. However, the downside is that pollution attacks become possible. A pollution attack occurs when a malicious (or faulty) node mixes invalid linear combinations of blocks into the network. These polluted blocks then get mixed with valid linear combinations and go undetected by honest intermediate nodes. The attack is detected only when the receiver is unable to reconstruct the original file, e.g. the reconstructed file hash does not match the original file hash. At this point, the entire file must be retransmitted from the source (if still available). To protect from pollution, homomorphic signatures (which are preserved through linear combinations) can be used. This provides non-repudiation and the ability to track and find malicious nodes. The drawback of homomorphic signatures is the Homomorphic NC processing cost, two order of  magnitude higher than the Conventional NC mixing cost - a prohibitive proposition in heterogeneous MANETs that include smart phones \cite{5978945}. While there exist less costly alternatives for preventing pollution attacks, these solutions place  limitation on topologies, require loose clock synchronization on the order of 100ms, limit the hop count, require large field sizes, or demand that new public keys be generated per generation. Obviously, these requirements are not feasible in dynamic CB-MANETs. 

This leaves us with two NC alternatives. One option is to perform source only coding, whereby only the publisher performs network coding and signs all the blocks. Since only the source codes and signs, non-repudiation is provided whereby the integrity of the blocks and the linear combination used to generate the blocks can be verified. Thus, receivers can identify pollution attacks and blacklist the malicious source. Another approach is to allow certified intermediate nodes that have fully reassembled the file to perform coding. To provide non-repudiation, the certified  intermediate node also signs the regenerated blocks in addition to the originator. In both these cases (source coding and full cache coding), non-repudiation is provided and thus downstream nodes are protected from untraceable pollution attacks.

The remainder of this paper discusses our system and the alternatives to unrestricted coding. We then perform a throughput comparison of the three options: unrestricted coding, full cache only coding, and finally source only coding.

\section{Network Coding Background} (TODO to be revised later)
We first begin with a brief overview of network coding \cite{ho2006random,fragouli2006network}. We start with a node which wishes to disseminate a file F which we call the source node. The source node first transforms F into a set of \textit{m} vectors $\mathbf{v_1, ..., v_m}$ in an n-dimensional vector space over a finite field $\mathbb{GF}_p$ where p is a prime number. These vectors are then linearly combined by drawing from the finite field $\mathbb{GF}_p$ an encoding coefficient $\mathbf{e_i}$ to linearly combine with the vector to create \textit{m} coded blocks $\mathbf{b_1, ..., b_m}$. The set of these coeffecients then forms the encoding vector \textbf{e} which [$\mathbf{e_1,...e_n}]$.

To reconstruct the file, a node simply must recover enough linearly independent coded blocks to be able to perform Matrix Inversion. First, we take the transpose of the received vectors such that $\mathbf{E^T}$ = [$\mathbf{{e^T}_{1}, ..., {e^T}_{n}}$], $\mathbf{B^T}$ = [$\mathbf{{b^T}_{1},...,{b^T}_{n}}$] and $\mathbf{V^T}$ = [$\mathbf{{v^T}_{1},...,{v^T}_{n}}$]. Then we take $\mathbf{E}^{-1}\mathbf{B}$ which will then reconstruct all the original blocks in the file.

\section{Content-Based Coding}
In content-based networks, addressing shifts from host-based addressing to content-based addressing. Each transmission unit (content block) is uniquely identified. In dynamic, intermittent  networks bandwidth is scarce. However, storage is becoming increasingly cheaper. The trend of increasingly cheap storage suggests to embrace the delay tolerant network philosophy of compensating for intermittent connectivity with, intermediate node caches. In case of popular files, this allows requestors to download from multiple caches even when the origin is unreachable.

In intermittent networks, the following challenges affect file dissemination:

\begin{itemize}
  \item \emph{Last coupon problem}: Teams may form and split frequently, thus a file must be transmitted (and can be retrieved from caches) in a piecemeal fashion. Thus, pieces are received out of order. This makes it difficult for the requestor to reliably reconstruct a file.
  \item \emph{Lack of end to end connectivity}: Hop by hop transmissions are required, with nodes acting as partial caches. Requestors must wait for the next contact opportunity to resume transmission.
  \item \emph{Partial caches}: Various nodes contain different parts of a file.
  \item \emph{Busy caches}: A requestor may find out that a cache which has the required pieces is busy serving other requestors. This causes the requestor to either wait for the next transmission opportunity or must locate another cache.
\end{itemize}

Content coding can help achieve reliable dissemination even when network partitions and severe disruptions occur and can address each of the above challenges. In particular:

\begin{itemize}
  \item \emph{Dispenses With Last Coupon Problem}: By using content coding, the last coupon problem is eliminated since with high probability each coded block received is innovative (i.e., helpful) and can be used to reconstruct the file. Thus, the throughput will be higher with content coding compared to not coding. 
\item \emph{Overcoming Intermittent Connectivity}: Since transmissions are connectionless and hop-by-hop, we cache blocks at intermediate nodes. A requestor can then ask nearby caches for network coded blocks. The neighbors pull coded blocks from their cache and either transmit as they are or mix them and transmit new coded blocks.
  \item \emph{Leverage Partial Caches}: Intermediate nodes cache partial files with innovative blocks. Since each block is helpful, nodes can make efficient use of limited connectivity by transmitting arbitrary innovative blocks.
   \item \emph{Parallel Cache Download}: When a requestor finds a nearby cache busy to answer requests, it can ask other nearby caches for blocks since each network coded block is as helpful as any other.
\end{itemize}

\section{Full Cache Coding}
In CB-MANETs, files are opportunistically cached at mobile nodes to favor future file requests. Files can be downloaded in parallel from multiple caches to make downloads reliable and fast. Network coding across parallel caches further improves the throughput. However, in intermittent connectivity, caches may often be partial. Thus, these caches cannot be signed since the signature implies that the intermediate node has received the full file, has verified the signature and has replaced in each block the originator signature with its own.  Note that an intermediate cache can reconstruct the file from contributions from different caches which is the traditional unrestricted coding. One may then state that the full cache strategy is like the unrestricted strategy in the following manner. As soon as an intermediate node decides to mix, it must fully reassemble the file and verify integrity before it reissues newly mixed packets. As we shall see, this intermediate full cache mixing can improve performance considerably as compared to source only coding.

The above implies that  the full cache strategy must be network coding aware. The question is whether a node should fully cache and decode/recode before forwarding (and signing) or should just forward the blocks as it receives them, no signature required.  There is a trade off between reassembly delay (-) and improved orthogonality (ie. linear independence) of the packets (+). We will show that in some cases we  can achieve higher throughput if we wait for the full cache. 

\subsection{Pollution Protection With Full Caches}
There are two types of pollution attacks to consider. The first is whereby a malicious node mixes and corrupts the coefficients such that downstream nodes are never able to successfully decode. The second attack occurs when blocks are polluted in such a way that downstream nodes are able to decode; yet, the reconstructed file's signature does not match the original file signature.

By pollution protection with full caches only, we mean the following. The file is first signed by the source, thus providing authentication, integrity, and non-repudiation. The signature can be saved in the header of the payload. Once an intermediate cache receives the full file, it verifies the source signature. Once the source signature has been validated, only then does the intermediate cache now assume responsibility for the integrity and non-repudiation. The file is random linearly network coded, and each block is signed by the cache owner (the intermediate node). Signing each block provides non-repudiation. By non-repudiation, we mean the following. If coded blocks from cache A cannot be decoded in spite of the collection of a full rank set, cache A is black-listed and avoided in the future (or inspected for faulty software). Now, with source signatures when the file is published and with intermediate node signatures after full cache reconstruction, the system is fully protected from pollution attacks.

Thus, to recover from either form of pollution attack, a receiver takes the following actions. Suppose a node receives from N caches and cannot decode. The receiver then requests blocks from a single cache at a time. The receiver must try to decode data from one cache at a time in order to isolate the faulty cache. The cache that provides an un-decodable stream or faulty signature  is the polluter and must be investigated.

\section{System Description}

The CB-MANET system used to evaluate the different NC pollution protection strategies 
subsumes both the family of peer-to-peer content dissemination network (e.g. Haggle[]) as well as the family of Content-Based Networks in which all blocks are uniquely identifiable (e.g. NDN []). To improve the delivery of content, we segment a large file into blocks. The transmissions are performed in the unit of data blocks and all blocks are named as filename/blockID. In the case of network coding, the blocks of a file are encoded as coded blocks and the block IDs are randomly generated. To leverage the wireless broadcast channel, all communications in this system are broadcast. 

\subsection{CB-MANET System Operation}

The basic operation of our system is as follows. Three messages are periodically broadcast at each node: interest, request, and cache summary. All three messages are represented in the form of bloom filters. The interest represents a collection of file names a node itself wants, and may be opportunistically disseminated over multiple hops when the bandwidth is sufficient. Relays have full control on when and which interests should be propagated. The request represents a collection of file the node is willing to receive at the time the request is sent. Note that the interest is separate from the request so that the node has the right to decide what contents to request based on the volume of interests it receive, the network condition, and its local content prioritization policy. Requests are broadcast only one hop to retrieve available contents from neighbors. To assist the prioritization and compactness of requests, nodes also periodically broadcast their cache summaries (in chunks or files, depending on the completeness of the data at this node.) The cache summaries are leveraged by neighbor nodes to decide which files/chunks to send. Cache summaries are useful in terms of reducing bandwidth waste, as nodes may blindly push redundant file/chunk based on a request for a large file.  Additionally, the nodes also update neighbors\' cache summaries based on the control messages and data communication they hear.

Data transmissions may be triggered when a new request comes in or when a new data is received. When a new request comes in, the node examines the request with the data it currently has and the requesters' cache summary, and initiates data transmissions for the data that matches the requester\'s request. A three-way handshake procedure is assciated with each data block transmission to eliminate redundant transmission in the broadcast network. For each data block, the sending node first sends a Request-To-Send-Block (RTSB), which carries the block name, to the target node. Upon receiving an RTSB, the target node replies a RTSB-Reply, which may accept or reject the block. If the block is rejected, a reject code is carried in RTSB-Reply to indicate one of the three reasons to reject: (1) The block is already received (2) The file is already received (3) The block is being sent by other neighbors. The data is then transmitted if accepted. Once the target node receives the data, it acknowledges by an ACK. Note that all neighbors of the target node may update their cache summaries based on the broadcast RTSB-Reply and ACK. 

When a data block is received, the receiver may propagate the data back to its original requester(s) by checking all requests it received from neighbors. If matches are found, the receiver (now becomes the sending node) starts another three-way handshake to deliver the block. In this way, the file or blocks are delivered back to the original requestors via the trail of breadcrumbs in a multi-hop environment. 

\section{Model}

\begin{figure}[t]
\centering
\includegraphics[width=0.2\textwidth]{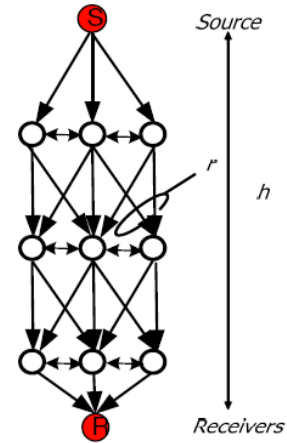}
\caption{1-3-3-1 corridor model \cite{Oh:2009:RMR:1702135.1702167}. There is a single node which broadcasts at the top. A single receiver subscribes to all files.}
\label{fig:static}
\end{figure}

Consider the corridor model that used in Oh and Gerla's 2009 paper \cite{Oh:2009:RMR:1702135.1702167}. The model depict two possible multipath configurations in a MANET, with perfectly disjoint and highly interfering paths respectively. The reality will be "in the middle", so we will study both cases and argue about average behavior.

In our case, the origin of the file is the node S. Node R has issued an interest for the file. The interest has traced several paths (as shown in Fig (a) and (b)). The file is splits into blocks which are broadcasted on the mesh (or braid) created by the interests. The broadcast mode precludes MAC layer ACKs so, no loss detection and retransmissions.

\subsection{No Coding}
	 	 	
First we consider the non coded file transmission. Each packet is triplicated by broadcast. Thus, 3 copies travel along the braid. With some probability a packet will be lost and the file cannot be properly received. Note that the probability of packet loss is much higher in (a) than in (b), because of greater (b) redundancy. However, (b) is slower than (a). More precisely, (b) is three times slower than (a), so in principle we could improve redundancy in (a) by transmitting each block 3 times. In general, with jamming and mobility, this non coding strategy becomes unacceptable.

\subsection{Unrestricted Coding}
	 	 	
 Consider first the braid (a). If packets are transmitted one at the time as soon as the get in, no mixing is possible and the performance is the same as for the non coded file.
In order to benefit from mixing we must accumulate at least 2 blocks and mix them to generate new mixed blocks. In this case, if a block is lost on a strand of the braid, the next coded block will allow recovery. The more blocks we accumulate in a queue, the more losses we can recover. Naturally, when the blocks are merged at the end, the triple redundancy of the three strands also comes to help. Next, consider the braid (b). In this case, because of the interconnection between the paths, two or three blocks are accumulated in each queue at each stage. These blocks can be mixed and will allow the recovery of the lost blocks. In summary, to exploit NC, we must ACCUMULATE AND MIX at intermediate nodes, at the expense of a few block delays. If blocks are not mixed, the performance is the same as for no coding

\subsection{Full Cache Only Coding}
	 	 	
Suppose the blocks are coded at the source and sent out in the network, ie they are broadcast on the braid. If there is no intermediate node mixing, the performance is the same as for uncoded file, i.e. potentially very bad. To improve the performance, the top three nodes in the corridor will assemble the file, while transmitting to the nodes below.

Once the top nodes have assembled the files, the mix them again and they transmit as many blocks as necessary to compensate for the lost blocks. In this particular case, the same outcome is achieved by transmitting additional blocks from the source. To guarantee that no blocks are repeated, each block is mixed (from the blocks in the cache) at the time it is transmitted. Loss recovery is very good. In fact as good as in the intermediate mixing case. Suppose ALL the blocks of the first broadcast are lost. Once the three caches start emptying their blocks, just 1/3 of blocks from each cache will suffice to recover the entire file. This is an interesting observation. If instead of building the three caches, we just relied on the source to transmit more coded blocks, we would take three times more to recover.

\subsection{Pre-Existing Caches}
	 	 	
Suppose now that there are several preexisting caches and that the interest query reaches multiple caches. If the caches are full caches, we can assume that they will be remixed as they are transmitted, so diversity is guaranteed and the recovery from loss very good. If the caches are partial caches and remixing is allowed, the mixing can be helpful if the caches are overlapped in the vector space. If the caches are disjoint subsets, as would be the case of a UAV spewing out the full coded file and 3 soldiers getting each a different 1/3 of the file, the mixing would not help much when they join and try to assemble the full file. Mixing will help if there is significant overlap in the files, as it will increase diversity. Anyway, as we saw earlier, if the soldier coded files already have built in diversity, further mixing at intermediate nodes will not help much. In summary, even with partial caches, the intermediate node mixing in general does not buy much.

\subsection{When does intermediate mixing help?}
	 	 	
Intuition suggests that intermediate node mixing helps if there is a sudden upsurge of losses and jamming that requires added redundancy. Suppose that the file is transmitted by the source in encoded fashion. Everything is fine for a few wireless hops and no copies are made nor mixing occurs at intermediate nodes. Until at some depth of the network intense jamming occurs that causes, say 90
With the caching strategy, one solution is actually possible if mixing upon full cache is allowed, namely reassembling the entire file at the upstream nodes and releasing freshly mixed blocks thereafter.  

\subsection{Hypothesis}

We have the following hypothesis: (1) intermediate full cache re-encoding provides better performance than source only network coding and (2) intermediate mixing is after all not very critical if diverse caches (full or not full) can be tapped in the network.

Our explanation is as follows:

\begin{itemize}

\item \emph{Explanation of hypothesis 1} (i.e.,  full cache re-encoding is better than source only coding). Suppose a receiver receives packets from two streams. If the streams have been independently coded by the upstream caches, the probability of getting a full rank set is higher than if the neighbors provide two identical sets.

\item \emph{Explanation of hypothesis 2} (i.e., full cache re-encoding provides comparable performance to unrestricted coding). In our scenarios, the main advantage of unrestricted coding is that even partial caches and, in the limited, nodes with two packets can mix thus creating more diversity. Also, unrestricted coding incurs less latency since it need not reassemble the entire file before mixing. However, unrestricted coding must accumulate and mix. Even in the extreme conditions when all else (but intermediate mixing) fails, the ability to create a full cache at node upstream of the critical section would save the day.
\end{itemize}

\section{Evaluation}
We evaluated the throughput of unrestricted coding, full-cache coding, source-only coding, and finally no coding. We examine both a static topology using the corridor model \cite{Oh:2009:RMR:1702135.1702167} with varying levels of packet loss and a mobile model using random waypoint. For simplicity, we evaluate a single generation, though our results are generalizable to multiple generations as our technique is not bound to generations.

\subsection{Evaluation Description}
Our test scenarios utilize multiple publishers disseminating using broadcast. The network is intermittent due to interference and induced packet loss, and in the dynamic case mobility. Many caches are partial; however, a receiver can download from multiple caches in parallel. Due to using broadcast there are no retransmissions. Redundancy is provided by multiple paths. After a time, the decoder discards the file that cannot be decoded. In both scenarios we perform our evaluation using Qualnet 6.1. The radio range is about 70 meters.

\subsection{Results}

\begin{figure}[t]
\includegraphics[width=0.5\textwidth]{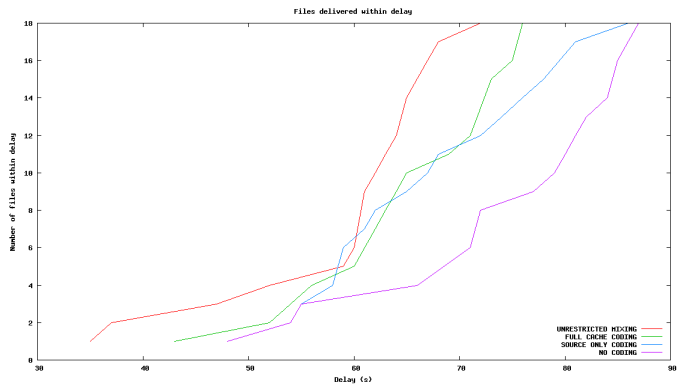}
\caption{Corridor model with 30\% packet loss. Single publisher and single downstream receiver with partial and full intermediate caches.}
\label{fig:static}
\end{figure}

\begin{figure}[t]
\includegraphics[width=0.5\textwidth]{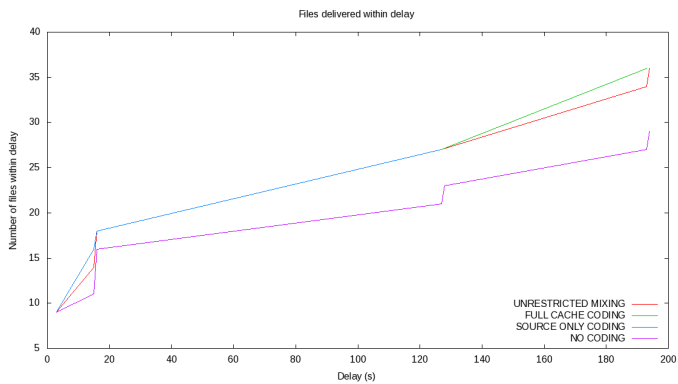}
\caption{10 node mobility with 3 publishers and 7 receivers.}
\label{fig:mobile}
\end{figure}

In our static scenario as seen in Figure ~\ref{fig:static}, we observe that as expected, full cache coding competes with unrestricted coding. This is due to the ability for the intermediate nodes to be able to quickly reconstruct the original file in parallel from the multiple publishers. The intermediate nodes act as seeds to propagate the file to the remaining receivers. Source only coding delivers files quicker than no coding. However, due to the bandwidth and processing overhead of network coding due to the encoding vector, source only coding is not able to as efficiently utilize the parallel caches as well as full cache coding.

During mobility as seen in Figure ~\ref{fig:mobile}, full cache coding performs as well as unrestricted mixing. As we observed in our hypothesis, unrestricted mixing gains its power when it is able to accumulate and mix new innovative blocks. During mobility, unrestricted mixing is not able to accumulate enough new innovative blocks, where the partial cache outperforms simply forwarding blocks. In the case of the full cache, unrestricted coding generates innovative blocks; however, so does full cache coding. This leads to the equivalent throughput.

\section{Related Work}

\subsection{Hybrid Solution}
Oh and Gerla showed that it is sufficient in a MANET for a small fraction of nodes to use homomorphic signatures with unrestricted network coding, while the other nodes simply forward \cite{5437114}. This is useful in heterogeneous radio scenarios with powerful laptops and light smart phones internetworked in the battlefield. Untrusted nodes are only able to forward blocks; thus, signatures are preserved and pollution attacks are prevented. Only trusted nodes are able to code and append a secure "digest" so that downstream nodes can verify the digest and discard polluted blocks.

\subsection{Network Coding Protocols}
CodeTorrent and CodeCast has studied network coding in MANETs whereby coded blocks are broadcasted and mixed at intermediate \cite{Lee:2006:CTC:1161252.1161254,4015713}. By exploiting partial caches, unrestricted coding is able to greatly decrease the delay required to deliver files.

\subsection{Security}
Homomorphic cryptography is computationally expensive and on the order of 2 times more expensive than unrestricted coding \cite{5978945}. This makes homomorphic cryptography infeasible for mobile devices such as smartphones.

More practical approaches for wireless networks have been proposed which utilize checksums \cite{dong2009practical}. However, these approaches require the receiver to establish loose time synchronization with the sender. Additionally, attacker identification requires joint cooperation between the receiver and source. Both of these constraints are difficult if not impossible to achieve in CB-MANETs and DTN type environments.

\section{Conclusion}
In conclusion, we have compared full cache coding with unrestricted coding in CB-MANETs and have shown that full cache coding competes or does as well as unrestricted coding. Our hypothesis that full cache coding is competes with unrestricted coding is true for both the static and mobile case. We find that unrestricted coding is dominated by intermittent connectivity; thus, limited in the ability to accumulate and mix. Our strategy of full cache coding which simply forwards blocks and waiting until only a full cache to re-encode, enables intermediate nodes to have comparable throughput to unrestricted coding.

%
\bibliographystyle{abbrv}
\bibliography{sigproc}  
%
%
\balancecolumns
\end{document}